# Gravity Well Echo Chamber Modeling With An LLM-Based Confirmation Bias Model

Joseph Jackson, Georgiy Lapin, Jeremy E. Thompson

# Abstract

An unfortunate byproduct of the recent boom in generative artificial intelligence has been a consequent boom in the spread of misinformation online. Through the widespread adoption of this new technology, many bad actors have sprung up, looking to use it for their own self-interest. This trend is most dangerous in political and other highly-stratified spheres because the harsh division of opinion is a further motivator for changing minds, and misinformation. Artificial intelligence, being effective for this use case, can be used as a way to disrupt the power balance of differing ideologies and cause a dramatic spike in misinformation used to push one worldview over another.

Our work expands upon existing gravity-well based online echo chamber models by including confirmation bias and improved metrics of ideological distance between a piece of content and a response to it. We use the model to identify echo chambers—which are known to harbor misinformation—with stronger pulls into the echo chamber on more confirmation-bias-prone users.

A fact integral to the spread of misinformation is the underlying presence of human confirmation bias whereby people seek information supporting existing beliefs, and reject any information opposing said beliefs. This variable was primarily introduced into the model because confirmation bias is ever-present both in the online and physical worlds, making many opinions more strongly held and, theoretically, making echo chambers more prevalent and powerful.

We propose a method for modeling confirmation bias and integrating it into the pre-existing gravity well echo chamber model as a way to best flag existing echo chambers online, the process of which slows the spread of misinformation by identifying its largest hubs. This confirmation bias variable is calculated user-by-user, incorporating repeated comparisons between the given user's post history and responses to differently opinionated posts from other users.

By performing a special calculation on these values, we arrived at a proper confirmation-bias-integrated version of the original gravity well model. Running this new model creates updated health markers showing a community's echo-chamber susceptibility. This model was validated on a set of nineteen subreddits from the Reddit social media platform. The ultimate contribution of this research is an improved method for recognizing fake news through detecting echo chambers in social media groups, helping mitigate the spread of misinformation by identifying its most common breeding ground.

# Table of Contents

# 1. Introduction

Online misinformation has become a serious concern as the recent development of generative artificial intelligence tools now allows for the rapid creation of false or misleading content. A big part of this issue lies in the formation of online echo chambers, where communities form around shared beliefs and opinions, reinforcing them while excluding differing viewpoints.[2] These echo chambers have been studied using the concept of gravity wells, where the "gravity" of group beliefs pulls users in, similar to how massive objects like stars or planets influence their surroundings in physics.[1]

However, the existing models that explain these echo chambers do not account for a major psychological factor: human confirmation bias.[7,8] Confirmation bias refers to the tendency of people to seek out and believe information that supports their existing views while ignoring or rejecting anything that contradicts them.[6]

In our work, we enhance the existing model by including confirmation bias and improving the way we measure how closely a user's beliefs align with the content they interact with. We hypothesize that adding a dynamic, user-specific confirmation bias parameter will improve the model's accuracy in identifying echo chambers. We also propose a more representative distance metric by using the top content from each subreddit, rather than just metadata or descriptions. We expect the enhanced model to align more closely with human understanding of echo chamber behavior, especially in edge cases.

## 2. Background

The spread of misinformation is significantly amplified in online environments through the formation of echo chambers—isolated communities where members primarily encounter opinions that reinforce their existing beliefs.[4,5] These structures not only filter out dissenting perspectives but also amplify partisan narratives. Prior studies have demonstrated the risks associated with these feedback loops, especially as they pertain to democratic discourse and decision-making.[2,3] To model the emergence and dynamics of such echo chambers, Thompson introduced a novel approach based on a gravity well simulation.[7] Drawing on analogies from astrophysics, the model conceptualizes ideological subgroups as mass-generating bodies in a shared opinion space.

Users are attracted to these subgroups based on several parameters:

- $m_{\text{subgroup}}$, representing the "mass" or size of the online community;

- $m_{\text{user}}$, originally a constant (set to 1) representing user confirmation bias strength;

- $TM$ (Technology Modifier), accounting for platform-level influence (e.g., Reddit vs. X);

- $TSM$ (Topic Source Modifier), differentiating content-specific traction within platforms;

- $d$, a BERT-based semantic distance metric measuring alignment between user and group ideology.[10]

These parameters are related in a gravity-well-resembling equation:

$$F_w \propto \frac{m_{\text{subgroup}} \times m_{\text{user}} \times \text{TM}}{d^2} \times \text{TSM}$$

*Eq.1 : Echo-chamber-modeling equation.*

The metaphor of gravity was effective in capturing how larger or more cohesive communities pull users into ideological alignment, forming *wells* that are difficult to escape. However, as Thompson acknowledges, the original model had a significant limitation: it did not dynamically account for confirmation bias. In this work, we build on and extend the previous model by integrating a dynamic, user-specific confirmation bias component. We also replace subreddit descriptions with aggregated high-engagement content to refine the ideological distance metric, allowing for a more content-driven measurement of user-group similarity. These modifications aim to produce a more realistic and granular model of online echo chambers, particularly suited to detecting misinformation-prone dynamics in modern digital platforms.

## 3. Problem

A key ingredient to the updated model is generative AI. While malicious actors have used generative AI to propagate misinformation on subreddits, generative AI also possesses significant potential for scalable investigation of numerous online communities, and for identifying potential echo chambers to mitigate their effect.

In particular, generative AI has been proven not only to generate misinformation, but to generate false explanations and justifications to propagate a belief in the generated misinformation, even amongst a critically-reasoning crowd.[9] This is clearly a significant problem which will only grow in prevalence online as generative AI becomes more powerful and more widely adopted for misinformation-propagating purposes. Given that even models from 2024 are able to convince critically-reasoning individuals to believe in falsehoods, it is not hard to assume that model improvements in the coming years will make genuine belief an easy goal for bad actors to accomplish. With the given proof that this misinformation is more successful at convincing people than traditional, human-generated misinformation, it is evident that a solution, or at minimum a quelling of this issue, must take place in order for the future of online and physical society to not be corrupted by widely-accepted misinformation.

# 4. Methodology

**Confirmation Bias Definition**

Confirmation bias is defined as the natural human tendency to select for information supporting existing beliefs while discrediting contradictory evidence. Based on this definition, if confirmation bias were to have a score, it would increase when a user either (1) intentionally validates a pre-formed opinion or (2) avoids contradictory information. Conversely, the confirmation bias score would decrease in the opposite scenarios: when a user (3) selects for contradictory opinions or (4) discredits existing beliefs.

**Modeling Confirmation Bias**

These four aforementioned conditions relate a user's sought-out information to their existing beliefs. The former is represented by new parent comments a user replies to, while the latter is represented by the historical parent comments. To clarify whether each parent comment is sought-out or discredited by the user, the model incorporates generative AI to determine the user's stance towards the parent comment. This support function is defined by its system prompt:

You'll receive:

- message: a comment in a Reddit thread

- parent: message's direct parent

- ancestors: the thread's previous post/comments (in chronological order, may be incomplete)

Evaluate how much the message supports the parent message. Ancestors provide contextualization.

Output format:

-1: vehemently opposes parent

-0.5: opposes parent with restraint

0: neutral

0.5: supports parent with restraint

1: passionately supports parent

[formatting details redacted]

*Fig.1 : System prompt for determining support.*

*The system prompt provides the message, its parent, and ancestors for context. The model asks the AI to quantify the message's support for its direct parent. The output ranges from -1 (vehement opposition) to 1 (passionate support). The support function's output is divided into intervals to distinguish categories for human–AI comparison. Altogether, the support function lets the model track what comments a user seeks and their relationship to them.*

Upon discovering the user's point of view towards the current parent comment, the model connects the current parent comment to historical parent comments. To determine their relationship, the model incorporates a new function called "alignment," which measures the alignment of the parent comments' opinions. For the old parent comment as well, the support function is run to determine the user's relationship to it. Each parent comment is compared to each parent comment before it. Eventually, all pairs are compared. Each comparison looks like this graph:

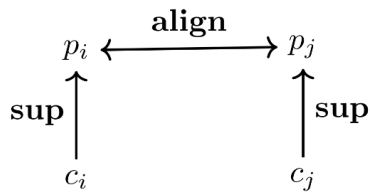

You'll receive two comment_contexts, each containing:

- message: comment in a Reddit thread

- parent: message's direct parent

- ancestors: the thread's previous post/comments (in chronological order, may be incomplete)

Evaluate how much the two messages' underlying opinions align, like an n-dimensional dot product.

Parents and ancestors provide contextualization.

Output format. The two messages' opinions...

-1: disagree

-0.5: might disagree

0: are independent, despite topic overlap

0.5: agree with restraint

1: ardently agree

Guidelines:

- Passionate opinions in similar topics may still be orthogonal: EVs make city streets way quieter. Mining for EV batteries wrecks ecosystems and exploits workers.

- Compare opinions, not facts. Opinions may be implicit, or expressed through tone.

[formatting details redacted]

*Fig.2 : System prompt for determining alignment. Again, intervals of 0.5 were chosen to provide clear boundaries between different scores, for evaluating human–AI results.*

*The system prompt provides the rest of the thread as context, and asks the AI to measure the alignment between the parent comments' opinions. The output can range from -1 (disagreement) to 1 (passionate agreement), split into intervals for human–AI comparison. Because the alignment function is more complex than the support function, further guidelines were provided.*

It was originally intended that distance be the metric relating parent comment opinions, but an exception was lack of differentiation between unrelated and opposing opinions. It was determined that alignment was more suitable as it could represent them as 0 and -1, respectively.

Dot products were ultimately chosen as the method of evaluating alignment because large language models have been successful representing language with vector spaces.[11]

The casework diagram illustrates the nuances of integer support and alignment.

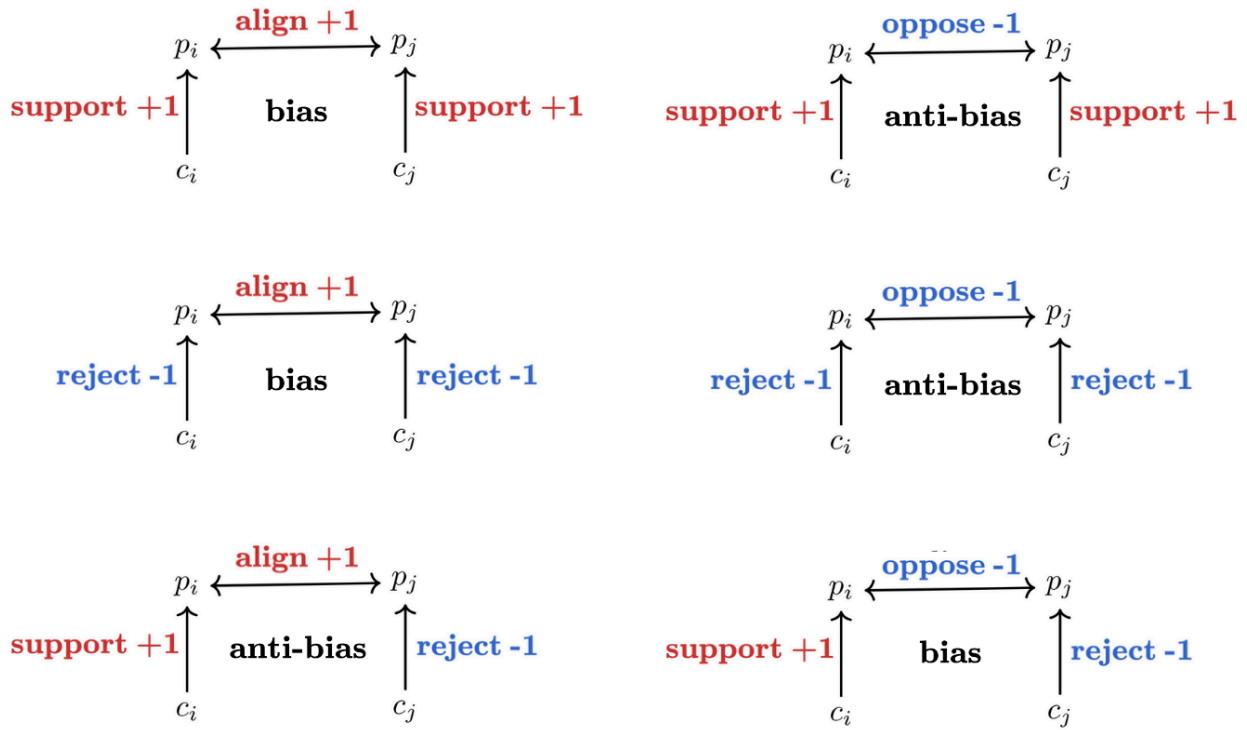

*Fig.3 : Casework logic for determining a comment pair's net confirmation bias contribution.*

*In the left column, parent comments have perfect alignment, so they express the same belief. Therefore, when the user holds the same stance for both comments in the pair, the user is seeking out information validating previous beliefs (confirmation bias). When the user holds different stances for the two comments between support and opposition, they are invalidating previous beliefs (not confirmation bias).*

*In the right column, the parent comments have perfect misalignment, so they express opposing beliefs. If the user has an equivalent strong partiality towards both comments, the user is*

*invalidating their previous beliefs (not confirmation bias). If the user shows strong support toward one comment and strong opposition toward the other, they are validating their previous beliefs (confirmation bias).*

The method discussed is formalized into the following expression. (Note: $m_a = m_{user}$)

$$m_{a,\text{unweighted}} = \frac{1}{\binom{n}{2}} \sum_{1 \le i < j \le n} c_i^{\sup} \otimes c_j^{\sup} \otimes \text{align}(p_i, p_j).$$

*Eq.2 : Confirmation-bias-modeling equation.*

Note: $\otimes$ is a known custom multiplication function. If two random variables uniformly distributed in [-1, 1] are multiplied, the result distribution naturally peaks near zero. $\otimes$ ensures the product remains uniformly distributed in [-1, 1]. $\otimes$ was chosen to spread the confirmation bias function's outputs.

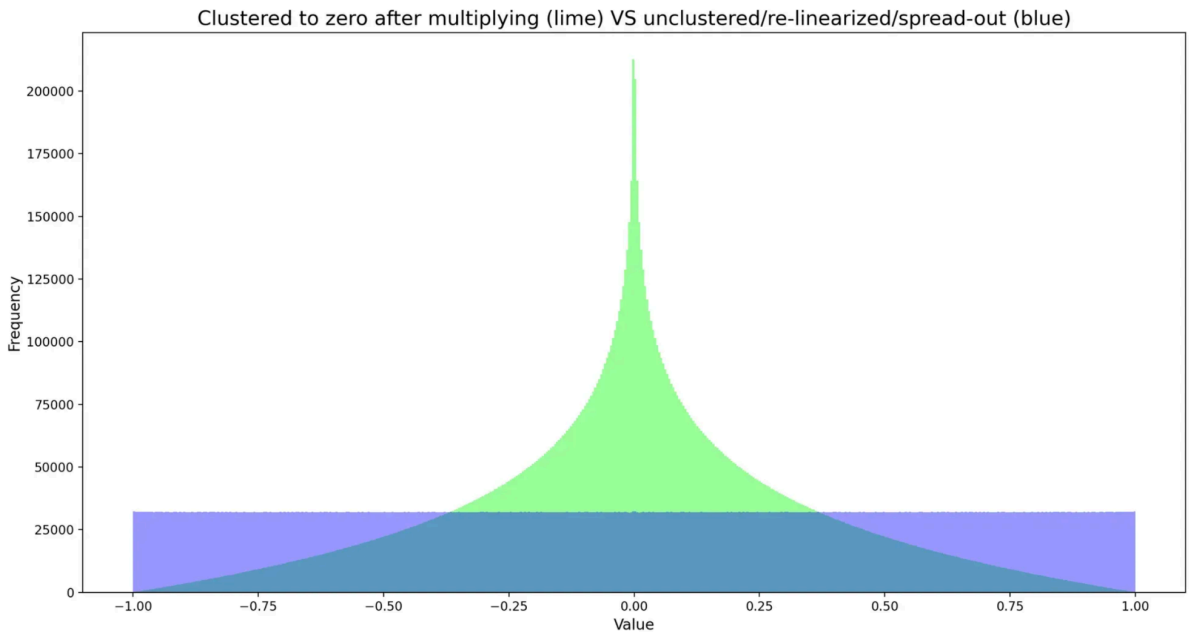

*Fig.4 : The blue-colored distribution represents the probabilities of a random variable uniformly distributed in [-1, 1]. The lime-colored distribution represents the probabilities of the product of two random variables.*

$$a \otimes b = \begin{cases} (ab)\,(1 - \ln|ab|), & ab \neq 0, \\ 0, & ab = 0. \end{cases}$$

*Eq.3 : An explicit formula for the custom multiplication function.*

Finally, substituting negative values into the original gravity well model would require a significant update to its formulation. Therefore, the final confirmation bias value is scaled from [-1, 1] into the range [0.5, 2], avoiding the issues that pertain to nonpositive values.

$$m_a = \begin{cases} 1.25 + 0.75\,m_{a,\text{unweighted}}, & n > 0 \\ 1, & n = 0 \end{cases}$$

*Eq.4 : Normalization of the bias value range.*

**Accuracy Evaluation**

For evaluation, we use exit order — the chronological order in which users become inactive in a subreddit — as a proxy for disengagement and, by extension, susceptibility to echo-chamber dynamics. Exit order is an indirect and noisy signal: inactivity can result from many processes (temporary breaks, account deletion, moderator bans, migration to other communities, or passive lurking) that do not necessarily reflect an ideological "exit." We used exit order because richer ground-truth signals (e.g., per-user view histories, private interaction logs, or platform-held engagement metrics) are not publicly available at scale due to platform data restrictions and privacy protections.

## 5. Results & Analysis

**Determining Human–AI Calibration**

The AI's results were validated against a human's to test synchronization. The human manually scored alignment and support on a random sample of comment pairs across the set of subreddits. Comparing humans to OpenAI's o4-mini-high model, the results are reflected in the two graphs.

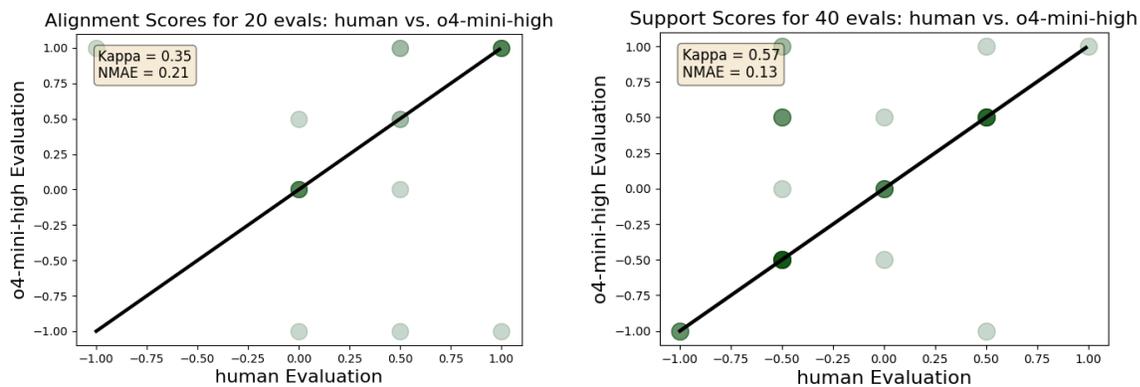

*Fig.5 : Human vs o4-mini-high scores, for 20 random alignment prompts and 40 random support prompts.*

For the comparison, OpenAI's generative model o4-mini-high was selected as it was then the most affordable, intelligent, and trustworthy model. The alignment function tests obtained a Quadratic-Weighted Kappa of 0.35, indicating fair agreement, and a Normalized Mean Absolute Error of 21%, indicating moderate deviation. Meanwhile, the support function tests resulted in a Quadratic-Weighted Kappa of 0.57, indicating moderate agreement, and a NMAE of 13%, indicating low deviation.

As tests show human–AI calibration is higher for support than for alignment, the concept of alignment is likely more complex than support. Overall, the calibration results are moderate, indicating that while humans and o4-mini-high mostly agree, there is room for improvement in the system prompt.

These experiments demonstrate that generative AI is a promising avenue for textual scoring in mathematical modeling in general, as while it is widely known it can be run at larger scales than humans, the agreement accuracy indicates the results are relevant as well.

**User Exit Order Comparison**

To determine how the original gravity well model would perform with confirmation bias, an exit order simulation was run. The original model predicts an exit order for each user, which is comparable to the actual exit order in which Reddit users became inactive. With confirmation bias integrated on samples of the subreddits, exit orders were simulated for eighteen subreddits, which were compared with real exit orders using Spearman's test.

| subreddit | n_common | spearman_rho | p_value |
|---|---|---|---|
| science | 34473 | -0.006313627858 | 0.879444531350 |
| cars | 21296 | -0.01670117555 | 0.992600367280 |
| travel | 12175 | 0.06979880594 | **0.000000000000** |
| math | 5951 | 0.02445433064 | **0.029623264415** |
| PoliticalDiscussion | 5709 | 0.05332218798 | **0.000027781926** |
| SandersForPresident | 5558 | 0.07124278034 | **0.000000052789** |
| hiking | 3309 | 0.02460121295 | 0.078558095500 |
| democrats | 3279 | 0.05883168584 | **0.000375176494** |
| Republican | 2170 | 0.0156215619 | 0.233512419400 |
| NeutralPolitics | 1822 | 0.03473209878 | 0.069174729100 |
| mlb | 1641 | 0.005481631386 | 0.412200037400 |
| flatearth | 1569 | 0.006885212062 | 0.392613688850 |
| progressive | 639 | -0.132709983 | 0.999614526556 |
| trump | 410 | 0.06694829862 | 0.088031117450 |
| SocialDemocracy | 157 | 0.02906175167 | 0.358934180950 |
| Freethought | 155 | 0.005748803681 | 0.471701557400 |
| AmericanPolitics | 141 | -0.04815776478 | 0.714670301350 |
| republicans | 84 | 0.03219464244 | 0.385631639750 |

*Table.1 : Spearman's tests between the confirmation bias-enhanced gravity well model's simulated exit orders and actual exit orders.*

For five of the eighteen sample subreddits, the p-value for Spearman's test is below 0.05.

Roughly one-third of the subreddit tests are significant, indicating likelihood that there exists a small but definite trace of correlation between the simulated and real exit orders. While the sample size is small, that the significant subreddits are larger overall indicates a correlation may exist between subreddit size and the p-value for Spearman's test, despite sampling.

# 6. Conclusions & Future Work

**Conclusions**

This research introduced a dynamic, user-specific confirmation bias variable into the gravity well model of echo chamber formation. By incorporating measures of user alignment with and responses to diverse viewpoints, the updated model provides a more psychologically realistic way to capture how individuals become entrenched in misinformation-prone communities. Tested across eighteen subreddits, the confirmation bias adjustment yielded modest but meaningful improvements in simulating exit order behavior compared to the original model. Unlike earlier approaches that relied primarily on community metadata or descriptions, our method integrates content-driven similarity and human cognitive tendencies, marking a step forward in echo chamber modeling.

While the gains observed in this study are incremental, they demonstrate that computational models of misinformation dynamics can be strengthened by grounding them in psychological realism. Even small improvements in echo chamber detection may aid researchers, platforms, and policymakers in identifying communities where misinformation thrives and in better understanding the mechanisms that keep users trapped within them. These results highlight the promise of generative AI not only as a source of misinformation but also as a scalable tool for modeling and detecting it.

**Future Work**

The confirmation bias model is likely best improved by reducing the time complexity. Currently, the time complexity is proportional to the square of the number of comments for each user, causing the model to be implausible to run for outlier users with many comments. The most promising solution is to weigh comment pairs over time, so that old comments are less relevant to new ones. Solving this time complexity issue will enable large-scale datasets to become viable.

In addition to time complexity, the confirmation bias and gravity-well model could also be strengthened if there were a more reliable real-world truth for comparative analysis, improving on exit order. Due to the inherent privacy of data such as views or likes, it is likely that such an improvement would entail large-scale human evaluation.

Another issue is filtering real humans from bots. The current confirmation bias and gravity well models ignore the bot existence, but as the amount of bots online increase and improve at blending in, the problem becomes more urgent. Large-scale Turing tests have been conducted such as Human or Not. However, bot filtering is a more challenging problem outside of the issue of confirmation bias.

Overall, it is desirable, despite the accuracy disruption due to bots, to have further similar research within social math modeling to identify and potentially combat the fake news that is propagated through social media, a pervasive issue in today's time.